\documentclass[12pt]{article}
\usepackage{amssymb, color}
\usepackage{bbm}
\usepackage{latexsym}
\usepackage{mathrsfs}
\usepackage{amsmath}

\usepackage{amsfonts}
\usepackage{indentfirst}
\usepackage{graphics}
\usepackage{bm}
\usepackage{natbib}
\usepackage{epsfig}
\usepackage[normalem]{ulem}
\usepackage{multirow}
\usepackage{siunitx}

\usepackage{bm}
\def\bbeta{{\bm\beta}}
\def\bx{{\bm x}}
\def\bX{{\bm X}}

\begin{document}
\thispagestyle{empty}
\parindent 4mm
\baselineskip 20pt
%\title{ A note }
%\author{}

%\begin{abstract}
%\end{abstract}

%\begin{center}
\title
{\large\bf
  Estimation for Dynamic and Static Panel Probit Models with
Large Individual Effects}
%\end{center}
\date{}

\author{\normalsize Wei Gao$^\star$ \qquad Wicher Bergsma$^*$  \qquad
Qiwei Yao$^{*,\dag}$\\[1ex]%}
%\address {
%\hspace{1em}
\normalsize
$^\star$Key Laboratory for Applied Statistics of MOE, School of Mathematics and Statistics, \\
\normalsize
Northeast Normal University, Changchun, Jilin 130024, China\\
%\hspace{1em}
\normalsize
%\address {
$^*$Department of Statistics, London School of Economics, London, UK\\
\normalsize
$^\dag$Guanghua School of Management, Peking University, Beijing, China}

\maketitle

\begin{abstract}
For discrete panel data, the dynamic relationship between successive observations is often of interest.
We consider a dynamic probit model for short panel data. A problem with estimating the dynamic parameter of interest is that the model contains a large number of nuisance parameters, one for each individual.
Heckman proposed to use maximum likelihood estimation of the dynamic parameter, which, however, does not perform well if the individual effects are large.
We suggest new estimators for the dynamic parameter, based on the assumption that the individual parameters are random and possibly large.
Theoretical properties of our estimators are derived and a simulation study shows they have some advantages compared to Heckman's estimator.

%Except that the variable given individual effects is logistically distributed, the individual effect is assumed to have some distributions with unknown parameters and then the maximum likelihood estimation is applied to estimate the dynamic pattern and unknown parameters. However, the analyzed results are sensitive to the chosen distributions of individual effects when observations for each individual are small, and this problem is more pronounced when the individual effects are large. In this paper, new estimation methods are proposed for dynamic and static probit models with short panel data, which are to some extent independent of individual effects. The properties of proposed estimators and the identification for parameters are given. Simulation studies show that the proposed statistics work well and the simulated results by our proposed methods are almost the same as those by Heckman's methods when individual effects are normally distributed.
\end{abstract}


\section{ Introduction}

\par
Short binary-valued time series in the presence of covariates are often
available in panel studies for which observations are taken on a panel of individuals
over a short time period. Dynamic probit regression is one of the
most frequently used  statistical models to analyse this type of data.
To set the scene, consider a panel of $n$ independently sampled individuals.
For each individual $i$, binary observations, denoted by $d_{i1}, \cdots, d_{iT}$,
are taken at time $1, \cdots, T$,  and the observations are assumed to satisfy the latent dynamic model:
\begin{equation} \label{LM}
d_{i1}=I(\tau_i+{\bx}_{i1}^{'}\bbeta+\epsilon_{i1}>0),\quad d_{it}=I(\tau_i+\gamma
d_{i, t-1}+{\bx}_{it}^{'}\bbeta+\epsilon_{it}>0)
\;\; {\rm for} \;\;
1< t \le T,
\end{equation}
where $I(\cdot)$ denotes the indicator function, $\{ \epsilon_{it}\}$  are independently and
identically distributed with mean $0$ and variance $1$, $\{{\bx}_{it}\}$ are $k\times 1$
covariate vectors, $\tau_i$ is an unknown intercept representing the $i$-th individual effect,
and the autoregressive coefficient $\gamma$ and
the regressive coefficient $\bbeta$ are unknown parameters which are
assumed to be the same for all individuals.  In (\ref{LM}), only the $d_{it}$ and $\bx_{it}$ are observable. The goal is often to estimate $\gamma$ and $\bbeta$ while the $\tau_i$ are treated as nuisance parameters.
As with most panel data, the number of individuals $n$ is large
while the length of observed time period $T$ is small. Therefore the asymptotic
approximations are often derived with $n\to \infty$ and $T$ fixed.
% It is assumed that $({\bx}_{i1},{\bx}_{i2})$ is independent of
% $(\epsilon_{i1},\epsilon_{i2})$. If the $\tau_i$ are random, then
% $({\bx}_{i1},{\bx}_{i2})$ is assumed independent of
% $(\tau_i,\epsilon_{i1},\epsilon_{i2})$.
%{\color{red} I have percent-out the condition on the independence to $\bx_{it}$.
%  Do we really need it, as they
%are observed and are effectively constants in the inference? -- QY}

Model (\ref{LM}) is a dynamic panel probit regression model, as the dynamic
dependence is reflected by the autoregressive parameter $\gamma$
which links $d_{it}$, i.e. the state at time $t$, to the state at time
$t-1$.  When $\gamma=0$, (\ref{LM}) reduces to a static panel probit regression, as
now $d_{it}$ is independent of $d_{i,t-1}, d_{i,t-2}, \cdots.$
Model (\ref{LM}) has been used for various applications in microeconomics by,
among others, Heckman (1978), Arellano and Honore (2001), and Hsiao (2003, Section 7.5).
 For example, Heckman (1978, 1980) used model (\ref{LM}) to reveal some interesting
dynamics in unemployment data: $d_{it}=0$ indicates that individual $i$ is unemployed at time $t$, and 1
otherwise, while the covariate $\bx_{it}$ stands for the factors
(such as age, education, family background etc) which may affect the
employment status.
These studies tried to provide statistical evidence to answer questions
such as: {\sl Does current unemployment cause future unemployment?}
If $\gamma > 0$ this indicates that being in employment at time $t$ increases the chances of being in employment at time $t+1$.

% The parameter $\gamma$ expresses the dynamic relationship between the
% previous state and the future state and is of considerable substantive
% interest. The state dependence for $\gamma\not=0$  has been termed as the
% real(or true) state dependence by Heckman (1978, 1980), which means that
% an individual who has experienced the event will behavior differently in
% the future compared with an otherwise identical individual who has not
% experienced the event.  For example, $d_{it}$ might represent employment
% status at time $t$, and the fact of being unemployed at $t=1$ is likely
% to result in a higher probability of being unemployed at time $t=2$ than
% can be explained by other covariates $({\bx}_{i1},{\bx}_{i2})$.  The
% state dependence for $\gamma=0$ has been termed as the spurious state
% dependence, in the sense that only temporally persistent unobservables
% determine the future value of $d$. The model has many applications in
% microeconomic data analysis.

Various estimation methods have been proposed for model (\ref{LM}).
By treating the individual effects $\tau_1, \cdots, \tau_n$ as
nuisance parameters or incidental parameters (Neyman and Scott, 1948),
Heckman (1980) adopted the maximum likelihood estimator of $\gamma$
as well as $\bbeta$ when $\epsilon_{it}$ are normally distributed.
Chamberlain (1980, 1985), Honore and Kyriazidou (2000), and Lancaster (2002)
considered the models with logistic distributed $\epsilon_{it}$. They proposed
a consistent estimator of $\gamma$ and derived its convergence rate.
Bartolucci and Farcomeni (2009) and Bartolucci and Nigro (2010) considered
some extended versions of dynamic logit models with
heterogeneity beyond those reflected by the covariates
in the models. A standard method to deal with incidental parameter problems%\sout{A distinguishing feature in handling many
%incidental parameters}
 is to use a conditional likelihood
to eliminate the incidental parameters by conditioning on sufficient
statistics for those parameters; see, e.g. Chamberlain (1980),
Bartolucci and Nigro (2010), and also Lancaster (2000).

An attractive alternative is to treat individual effects $\tau_i$
as random effects with prespecified priors.
But as far as we are aware, most literature on panel probit regression
taking this approach only deal with the static model (i.e. $\gamma=0$ in (\ref{LM})) only.
For example, Chamberlain (1980, 1985) discussed the maximum likelihood estimator for $\bbeta$
with a given prior distribution for $\tau_i$.
Arellano and Bonhomme (2009) showed that this estimator is robust
with respect to the choice of prior when $T$ is large.
% But for a short panel as the present
% case ($T=2$), different priors may lead to quite different estimators of
% $\gamma$ and so the prior needs to be carefully chosen. In most cases, we
% do not know how to choose a suitable prior.
Manski (1987) proposes maximum score methods to estimate $\bbeta$  when
the distribution of the errors is unknown and $\gamma$ is equal to zero
for model (\ref{LM}). Smoothed maximum score estimators were developed by
Horowitz (1992). See also Arellano (2003) for  a survey of static probit models.
% panel data for static models with individual effects is given by Arellano
% (2003).
% By introducing a quadratic exponential model, Bartolucci and
%Farcomeni (2009), Bartolucci and Nigro (2010) consider estimating
%problems for binary panel data.

In this paper, we propose new estimation methods for $\gamma$ and $\bbeta$
in model (\ref{LM}) with  $\epsilon_{it} \sim N(0, 1)$.
We treat $\tau_i$ as random effects but with an unspecified prior.
Our methods are designed for the cases when the individual effects $\tau_1,
\cdots, \tau_n$ are large while $T$ is small.
Note that when $\tau_i $ are large, there is an innate difficulty in estimating
$\gamma$ and $\bbeta$ as the outcome of the random event $\{ \tau_i + \gamma
d_{i, t-1} + \bx_{it}' \bbeta + \epsilon_{it}>0 \}$ may be dominated by
the value of $\tau_i$. In fact Heckman (1980) reported that the maximum
likelihood estimator for $\gamma$ behaved poorly when the variance of $\tau_i$
is large; see Table 4.2 in Heckman (1980).
Furthermore, our simulation results indicate that our methods work as well
as Heckman's (1980) method when the variance of $\tau_i$ are, for example, equal
to 1 and 4.

The rest of the paper is organised as follows: Section 2 presents the new
estimation methods together with their asymptotic properties. For the simplicity
of the presentation, we consider the case $T=2$ only, though the methods can
be extended to the cases with $T>2$. %{\color {red} GW: Do you agree?}.
Simulations are reported in Section 3 and an example  is analyzed in Section 4.
%Numerical illustration with both simulated and real {\color{red} ***IF AT ALL POSSIBLE***}
%data is reported in Section 4.
 Some technical proofs are relegated to the Appendix.

\section{Estimation methods}

We consider model (\ref{LM}) with $T=2$, namely
\begin{equation} \label{b1}
d_{i1}=I(\tau_i+ \bx_{i1}' \bbeta +\epsilon_{i1}>0), \quad
d_{i2}=I(\tau_i+ \gamma d_{i1}+ \bx_{i2}' \bbeta
+\epsilon_{i2}>0), \quad i=1, \cdots, n,
\end{equation}
where $\{\epsilon_{i1}\}$ and $\{ \epsilon_{i2}\}$ are independent and
$N(0, 1)$, and $\tau_i$ is independent of $\epsilon_{i1}$ and $\epsilon_{i2}$.
Furthermore, we assume that
 $\{\tau_i\}$ are
independent with a common density function $f(.)$ which satisfies condition C1 below.
\begin{quote}
\begin{description}
\item[C1]
The density function of $\tau_i$ admits the expression
\begin{equation}
f(x)=\frac{1}{\sigma_{\tau}}h(\frac{x-\mu_\tau}{\sigma_{\tau}}),\label {SU}
\end{equation}
where $h(\cdot)$ is a density function with mean 0 and variance 1, $h(x)$ is
continuous at $x= 0$, and $\mu_\tau $ and $\sigma_\tau>0$ are constants.
\end{description}
\end{quote}

We present below the new estimation methods for the three scenarios: (i) estimating
the autoregressive coefficient
$\gamma$ without covariates (i.e. $\bbeta=0$), (ii) estimating regressive coefficient
vector $\bbeta$ for the static model (i.e. $\gamma=0$), and (iii) estimating
$\gamma$ and $\bbeta$ together. All those methods are derived based on  some
asymptotic arguments when $\sigma_\tau \to \infty$, and therefore the methods
are particularly relevant when the individual effects are large.

\subsection{Estimation of $\gamma$ when $\bbeta=0$}

%\par
When $\bbeta=0$,  model (\ref{b1}) reduces to
\begin{equation}
d_{i1}=I(\tau_i+\epsilon_{i1}>0),\quad\;d_{i2}=I(\tau_i+\gamma
d_{i1}+\epsilon_{i2}>0),\hspace{6mm}i=1,\ldots,n. \label{LM2}
\end{equation}
As $\tau_i$, $\epsilon_{i1}$ and $\epsilon_{i2}$ are independent, and
$\epsilon_{i1}$ and $\epsilon_{i2}$ are $N(0,1)$, it holds that
\begin{align} \label{b2}
P\{d_{i1}=0,d_{i2}=0\} &=\int\Phi(-x)\Phi(-x)f(x)dx,\\ \nonumber
P\{d_{i1}=0,d_{i2}=1\} &=\int\Phi(-x)\Phi(x)f(x)dx,\\ \nonumber
	P\{d_{i1}=1,d_{i2}=0\} &=\int\Phi(x)\Phi(-x-\gamma)f(x)dx,\\ \nonumber
P\{d_{i1}=1,d_{i2} =1\}& =\int\Phi(x)\Phi(x+\gamma)f(x)dx,
\end{align}
where $\Phi$ is the standard normal distribution function, and $f(\cdot)$ is
the density function of $\tau_i$. We state in Proposition 1 below an asymptotic property
on the ratio of the two probabilities listed above, on which our new
estimation method for $\gamma$ is motivated. Its proof follows
immediately from Lemmas 1 and 2 in the Appendix.
% By the comments given by Heckman (1978), we can deduce the sign
% of $\gamma$ from the ratio of $P\{d_{i1}=1,d_{i2}=0\}$ and
% $P\{d_{i1}=0,d_{i2}=1\}$, which can be estimated by
\vskip1em
\par
{\bf Proposition 1.} Under condition C1,
it holds that
\begin{align}
\label{HE}
\lim_{\sigma_\tau\to \infty}
{ \frac{P\{d_{i1}=1,d_{i2}=0\}}{P\{d_{i1}=0,d_{i2}=1\}}} = &
\displaystyle\lim_{\sigma_\tau\rightarrow\infty}
\frac{\int\Phi(x)\Phi(-x-\gamma)f(x)dx}{\int\Phi(-x)\Phi(x)f(x)dx}\\
=& \frac{\int\Phi(x)\Phi(-x-\gamma)dx}
{\int\Phi(x)\Phi(-x)dx}= G(\gamma),
\nonumber
\end{align}
where
\begin{equation} \label{b3}
G(\gamma)=
-\sqrt{\pi}\gamma\Phi(-\frac{\gamma}{\sqrt{2}})+\exp\{-\frac{\gamma^2}{4}\}.
\end{equation}
\par
	\vskip1em

	\par
Proposition 1 above suggests the following estimator for $\gamma$:
\begin{equation}
\widehat{\gamma}=G^{-1}(\widehat W),\label{ES}
\end{equation}
where $G(\cdot)$ is given in (\ref{b3}), and
\begin{equation}
\widehat
W=\sum\limits_{i=1}^{n}I(d_{i1}=1,\;d_{i2}=0)
\Big/
\sum\limits_{i=1}^{n}I(d_{i1}=0,\;d_{i2}=1),\label{RO}
\end{equation}
i.e. $\widehat W$ is a plug-in estimator for the ratio of the two
probabilities on the left hand side of (\ref{HE}). The asymptotic properties of
$\widehat \gamma$ are stated in the theorem below. Put
\begin{equation} \label{b4}
\kappa_n = \big\{ \sum_{i=1}^n I(d_{i1}=0, \, d_{i2} =1 ) \big\}^{1/2},
\quad
\sigma^2=\frac{G(\gamma)+G^2(\gamma)}{[G^{'}(\gamma)]^2}=
\frac{G(\gamma)+G^2(\gamma)}{\pi\Phi^2(-\gamma/\sqrt{2})}.
\end{equation}

\vskip1em
{\bf Theorem 1.} Under condition C1, the following assertions holds.

(i) $ \lim_{\sigma_\tau\rightarrow\infty}
\lim_{n\rightarrow\infty}P\{|\hat{\gamma}-\gamma|\geq\eta\}=0$ for any $\eta>0$.

(ii) $ \lim_{n\to \infty} P\{ \kappa_n (\widehat \gamma - \gamma) \le x \} =
\Phi(x/\sigma)$ for any real number $x$, provided that the first
derivative of $h(\cdot)$ is continuous and $\sigma_\tau = a \sqrt{n}$ for
some constant $a$.
\vskip1em

{\bf Remark 1}. (i) Theorem 1(i) can be viewed as a version of consistency for
$\widehat \gamma$. Theorem 1(ii) indicates that $\widehat \gamma$ is asymptotically normal
if we restrict $\sigma_\tau = a \sqrt{n}$. Note that the convergence rate $\kappa_n$
defined in (\ref{b4}) admits
the  asymptotic relation:
the standard $\sqrt{n}$, as
\[
\kappa_n^2 = n f(\mu_\tau) \int \Phi(u) \Phi(-u) du + o_p(n/\sigma_\tau).
\]
See the proof of Theorem 1 in the Appendix.
Thus the larger the sample size $n$ is, the faster $\widehat \gamma$ converges to $\gamma$.

(ii) Only two out of the four probabilities in (\ref{b2}) are used in defining the
estimator $\widehat \gamma$. Indeed the observations with $(d_{i1}, d_{i2}) = (0, 0)$
or $(1, 1)$ are not utilized in (\ref{ES}). In fact when $\sigma_\tau$ is large,
those data provide little information on $\gamma$, as
\begin{align*}
& \lim_{\sigma_\tau\rightarrow\infty}P\{d_{i1}=0,d_{i2}=0\}
% =\lim_{\sigma_\tau\rightarrow\infty}\int\Phi(-x)\Phi(-x)f(x)dx\\
=\lim_{\sigma_\tau\rightarrow\infty}\int\Phi(-x)\Phi(-x)\frac{1}
{\sigma_{\tau}}h(\frac{x-\mu_\tau}{\sigma_{\tau}})dx\\
&= \lim_{\sigma_\tau\rightarrow\infty}\int\Phi(-\sigma_\tau t-\mu_\tau)\Phi(-\sigma_\tau t-\mu_\tau)h(t)dt
=H( 0),
\end{align*}
	and similarly
	\begin{eqnarray*}
	\lim_{\sigma_\tau\rightarrow\infty}P\{d_{i1}=1,d_{i2}=1\}
% &=&\lim_{\sigma_\tau\rightarrow\infty}\int\Phi(x)\Phi(x+\gamma)f(x)dx\\
% &=&\lim_{\sigma_\tau\rightarrow\infty}\int\Phi(x)\Phi(x+\gamma)\frac{1}{\sigma_{\tau}}h(\frac{x-\mu_\tau}{\sigma_{\tau}})dx\\
%	&=&\lim_{\sigma_\tau\rightarrow\infty}\int\Phi(\sigma_\tau t+\mu_\tau)\Phi(\sigma_\tau t+\mu_\tau+\gamma)h(t)dt\\
=1-H( 0).
\end{eqnarray*}
where $H(x)$ is cumulative distribution function of $h(x)$.
% This may be a reason that the maximum likelihood estimator of $\gamma$ performs badly for large $\sigma_\tau$ in the simulation study given by Heckman (1980). The variance of parameters will become larger when additional information has no direct connection with the parameters of interest.

\subsection{ Estimation of $\bbeta$ when $\gamma=0$}

\par
Let
$$
D_n=\{(d_{i1},d_{i2})^{'}:\;\;d_{i1}+d_{i2}=1\;\;\mbox{for}\;\;i=1,\cdots,n\}
$$
and denote the number of elements in $D_n$ by $m$. Without loss of generality, suppose that
$d_{i1}+d_{i2}=1$ for $i=1,\cdots,m$.

We find the conditional probability
\begin{eqnarray*}
&&P\{d_{i1}=1,d_{i2}=0|d_{i1}+d_{i2}=1,\;{\bx}_{i1},{\bx}_{i2}\}\\
&&=\frac{\int\Phi({\bx}_{i1}^{'}\bbeta+t)\Phi(-{\bx}_{i2}^{'}\bbeta-t)f(t)dt}
{\int\Phi({\bx}_{i1}^{'}\bbeta+t)\Phi(-{\bx}_{i2}^{'}\bbeta-t)f(t)dt+\int\Phi(-{\bx}_{i1}^{'}\bbeta-t)\Phi({\bx}_{i2}^{'}\bbeta+t)f(t)dt}.
\end{eqnarray*}
Under (\ref{SU}), we can similarly prove
$$
\lim_{\sigma_\tau\rightarrow\infty}P\{d_{i1}=1,d_{i2}=0|d_{i1}+d_{i2}=1,\;{\bx}_{i1},{\bx}_{i2}\}=\frac{G(({\bx}_{i2}-{\bx}_{i1})^{'}\bbeta)}
{G(({\bx}_{i2}-{\bx}_{i1})^{'}\bbeta)
+G(-({\bx}_{i2}-{\bx}_{i1})^{'}\bbeta)}.
$$
For sufficiently large $\sigma_\tau$,  we can replace the conditional likelihood of $\bbeta$ given $D_n$  by
\begin{equation}
L(\bbeta)=\prod_{i=1}^{m}p_{i}^{z_i}(1-p_i)^{1-z_i}\label{CL}
\end{equation}
where $z_i=I(d_{i1}=1,d_{i2}=0)$ and $1-z_i= I(d_{i1}=0,d_{i2}=1)$, and
\begin{equation}
p_i=\frac{G(({\bx}_{i2}-{\bx}_{i1})^{'}\bbeta)}{G(({\bx}_{i2}-{\bx}_{i1})^{'}\bbeta)
+G(-({\bx}_{i2}-{\bx}_{i1})^{'}\bbeta)}.\label{CP}
\end{equation}
\vskip1em
Note that $p_i=K(({\bx}_{i2}-{\bx}_{i1})^{'}\bbeta)$ for the monotone function $K$ defined as
$$
K(t)=\frac{G(t)}{G(t)+G(-t)},
$$
Hence,~(\ref{CP}) is a generalized linear model of the form
$$
K^{-1}(p_i)=({\bx}_{i2}-{\bx}_{i1})^{'}\bbeta.
$$
So iterative reweighted least squares methods for generalized Models given by McCullagh and Nelder (1989) can be applied to (\ref{CL}) to estimate the parameter $\bbeta$. Under some regularity conditions and $\sigma_\tau\longrightarrow\infty$, consistency of $\bbeta$ can be shown.

\subsection{ Simultaneous estimation of $\gamma$ and $\bbeta$}

As in Section 2.2, we have
$$
\lim_{\sigma_\tau\rightarrow\infty}P\{d_{i1}=1,d_{i2}=0|d_{i1}+d_{i2}=1,\;{\bx}_{i1},{\bx}_{i2}\}=
\frac{G(\gamma+({\bx}_{i2}-{\bx}_{i1})^{'}\bbeta)}{G(\gamma+({\bx}_{i2}-{\bx}_{i1})^{'}\bbeta)
+G(-({\bx}_{i2}-{\bx}_{i1})^{'}\bbeta)}.
$$
For large $\sigma_\tau$, we replace the conditional likelihood of $\gamma$ and $\bbeta$ given $D_n$ by
\begin{equation}
L(\bbeta)=\prod_{i=1}^{m}p_{i}^{z_i}(1-p_i)^{1-z_i}\label{CL1}
\end{equation}
where $z_i=I(\{d_{i1}=1,d_{i2}=0)$ and $1-z_i= I(d_{i1}=0,d_{i2}=1)$, and
\begin{equation}
p_i=\frac{G(\gamma+({\bx}_{i2}-{\bx}_{i1})^{'}\bbeta)}{G(\gamma+({\bx}_{i2}-{\bx}_{i1})^{'}\bbeta)
+G(-({\bx}_{i2}-{\bx}_{i1})^{'}\bbeta)}.\label{CP1}
\end{equation}

\par
Let
$$
\bX^{*}=\left({\bx}_{12}-{\bx}_{11}, {\bx}_{22}-{\bx}_{21},\cdots,{\bx}_{m2}-{\bx}_{m1}\right)
  $$

\par
{\bf Theorem 2.}  (\ref{CP1}) is identifiable for $\gamma$ and $\bbeta$ if the rank of $\bX^{*}$ is equal to $k$ (the dimension of ${\bx}_{2i}-{\bx}_{1i}$) and at least there exist $j$ and $1\leq s_1,\cdots,s_k\leq m$ such that
$$
{\bx}_{j2}-{\bx}_{j1}=a_1({\bx}_{s_1 2}-{\bx}_{s_1 1})+a_2({\bx}_{s_2 2}-{\bx}_{s_2 1})+\cdots+a_k({\bx}_{s_k 2}-{\bx}_{s_k 1})
$$
where $a_1,\cdots,a_k$ are non-positive real numbers.

\vskip1em

The conditions in Theorem 2 are sufficient and can be satisfied with probability close to $1$ for a large sample size $n$ if the covariate ${\bx}_{i2}-{\bx}_{i1}$ is a continuous variable and its covariance matrix is positive definite.
\vskip1em
{\bf Corollary.} Under the condition in Theorem 2, and with ${\bf 1}_{m}$ be the $m-$dimensional vector with all components $1$, the rank of $({\bf 1}_m,{\bX^{*}}^{'} )$ is $k+1$.

\vskip1em

From the Corollary, it seems that the identifiability condition relating to (\ref{CP1}) is stronger than that of linear models since that the rank of design matrix being equal to the number of parameters is sufficient for linear models to be identified.

\par

\section{Simulation study}

In this section, we use simulations to estimate the root mean squared errors (RMSEs) of the estimators proposed in Section 2.
In Table \ref{table1}, RMSEs of $\gamma$ in Model (\ref{LM2}) are given for different distributions of the individual effects.
In Table \ref{table2}, RMSEs of $\gamma$ and $\beta$ in Model (\ref{LM}) are given, with the $x_{i1}$ sampled from the standard normal distribution and $x_{i2}=x_{i1}+N(0,1)$; the individual effects are normally distributed with mean $0$ and variance $2$.
For normally distributed individual effects with mean $0$ and variance $\sigma^2$ in Model (\ref{LM}), Heckman (1980) has proposed the maximum likelihood estimation of the dynamic parameter $\gamma$ and $\sigma^2$. In Tables \ref{table3} and \ref{table4} the RMSE of our new estimator is compared with the RMSE of Heckman's estimator, in the former table for normally distributed individual effects and in the latter for individual effects with a mixture normal distribution.
We see that our estimator is comparable to Heckman's for normally distributed effects with moderate variance, but greatly outperforms it when individual effects are mixed normal distributions.

\begin{table}
\caption{Simulated RMSEs of the new estimator of the dynamic parameter $\gamma$ in Model (\ref{LM2}) (100 replications)}
\label{table1}\centering
\begin{tabular}{c  S[table-format=2.1] S[table-format=1.2] r S[table-format=1.2] c }

\hline\\
                     &                 &  \multicolumn{1}{r}{$n=1000$}       &          & \multicolumn{1}{l}{$n=5000$}                \\
     Distribution of the $\tau_i$  &    $\gamma$ & {RMSE}   &  \hspace*{10mm}Distribution of the $\tau_i$     &  {RMSE}  \\
 \hline
               &     -2    &               0.16    &                 &         0.21   \\
                &     -1.5  &              0.24      &             &           0.19   \\
                &     -1    &               0.23     &            &             0.14  \\
                &     -0.5  &              0.20    &             &             0.15   \\
   U(-3,3)     &     0     &             0.15      & U(-10,10)   &              0.13   \\
                &     0.5   &            0.21       &                &            0.31       \\
                &     1     &              0.18       &                 &              0.14       \\
                &     1.5   &               0.15     &                &              0.15  \\
                &     2     &               0.25      &                 &              0.18 \\
   \hline
                &     -2   &               0.30      &               &              0.16\\
                &     -1.5  &             0.15       &             &             0.19 \\
                &     -1    &               0.20     &              &               0.13  \\
                &     -0.5  &               0.15     &              &               0.12       \\
   N(0,4)      &     0     &               0.15      & N(0,25)        &              0.11  \\
                &     0.5   &              0.16       &                &              0.10       \\
                &     1     &             0.17      &                 &              0.11    \\
                &     1.5   &               0.18      &                &              0.12 \\
                &     2     &               0.23      &                 &               0.17  \\
                                      \hline

 \end{tabular}
\end{table}

\begin{table}
\caption{Simulated RMSE of new estimators of $\gamma$ and $\beta$ for Model (\ref{LM}) ($200$ replicates and $n=1000$)
\label{table2}}\centering
\begin{tabular}{  c c    c c c c c c}

\hline\\
  $\gamma$ &  $\beta$    & RMSE($\hat{\gamma}$) & RMSE($\hat{\beta}$) &$\gamma$ &  $\beta$    &  RMSE($\hat{\gamma}$ )& RMSE($\hat{\beta}$)\\
 \hline
-1&0&0.20&0.08 & 0&-1&0.20&0.15\\
-0.5&0&0.17&0.08&0&-0.5&0.18&0.10 \\
0&0&0.14&0.08\\
0.5&0&0.16&0.08&0&0.5&0.16&0.10\\
1&0 &0.16&0.09&0&1&0.19&0.13\\
\hline
-1&1&0.22&0.13&1&1&0.25&0.16\\
-0.5&0.5&0.19&0.10&0.5&0.5&0.15&0.09\\
0.5&-0.5&0.16&0.10&-0.5&-0.5&0.17&0.10\\
1&-1&0.22&0.18&-1&-1&0.24&0.13\\
\hline
\end{tabular}
\end{table}

%\begin{table}
%\caption{Simulated RMSE of new estimators of $\gamma$ and $\beta$ for Model (\ref{LM}) ($200$ replicates and $n=1000$)
%\label{table2}}\centering
%\begin{tabular}{  c c    c c c c c c}

%\hline\\
%  $\gamma$ &  $\beta$    &  $\hat{\gamma}$(se) & $\hat{\beta}$ (se)&$\gamma$ &  $\beta$    &  $\hat{\gamma}$(se) & $\hat{\beta}$(se)\\
% \hline
%-1&0&-0.97(0.20)&0.01(0.07) & 0&-1&-0.01(0.20)&-1.01(0.15)\\
%-0.5&0&-0.5(0.17)&-0.01(0.08)&0&-0.5&-0.01(0.18)&-0.50(0.10) \\
%0&0&-0.00(0.14)&0.01(0.08)\\
%0.5&0&0.49(0.16)&0.01(0.08)&0&0.5&0.00(0.16)&0.50(0.10)\\
%1&0 &1.00(0.16)&0.01(0.09)&0&1&0.03(0.18)&1.00(0.13)\\
%\hline
%-1&1&-0.94(0.21)&0.96(0.12)&1&1&1.06(0.24)&1.04(0.15)\\
%-0.5&0.5&-0.50(0.19)&0.50(0.10)&0.5&0.5&0.51(0.15)&0.51(0.09)\\
%0.5&-0.5&0.51(0.16)&-0.49(0.10)&-0.5&-0.5&-0.49(0.17)&-0.48(0.10)\\
%1&-1&1.03(0.22)&-1.05(0.17)&-1&-1&-0.96(0.24)&-0.96(0.12)\\
%\hline
%\end{tabular}
%\end{table}

\begin{table}
\caption{Comparison of RMSE of new estimator ($\hat\gamma_G$) and Heckman's estimator ($\hat\gamma_H$) for normally distributed individual effects ($200$ replicates for sample size $n=1000$). }\label{table3}\centering
\begin{tabular}{c c c c c}
\hline\\
  Distribution of the $\tau_i$      &$\gamma$& RMSE(${\hat\gamma}_G$)& RMSE(${\hat\gamma}_H$)&RMSE(${\hat\sigma}_H$)\\
        \hline\\
         &  -1   &0.16&0.13&0.13\\
         & -0.5  &0.14&0.11&0.12 \\
$N(0,1)$ &   0   &0.12&0.09&0.11\\
         & 0.5   &0.13&0.10&0.11\\
         &1      &0.13&0.10&0.12\\
         \hline\\
          & -1  &0.20&0.16&0.25 \\
         & -0.5 &0.18&0.15&0.21  \\
$N(0,4)$ &   0  &0.15&0.12&0.18 \\
         & 0.5  &0.17&0.14&0.26\\
         &1     &0.17&0.15&0.20 \\
         \hline \\
\end{tabular}\\
\end{table}

%\begin{table}
%\caption{Comparison of RMSE of new estimator ($\hat\gamma_G$) and Heckman's estimator ($\hat\gamma_H$) for normally distributed individual effects ($200$ replicates for sample size $n=1000$). %}\label{table3}\centering
%\begin{tabular}{c c c c c}
%\hline\\
%  individual effects      &$\gamma$& ${\hat\gamma}_G$& ${\hat\gamma}_H$&${\hat\sigma}_H$\\
%        \hline\\
%         &  -1   &-0.83(0.15)&-1.01(0.13)&1.04(0.12)\\
%         & -0.5  &-0.47(0.13)&-0.54(0.11)&1.06(0.10) \\
%$N(0,1)$ &   0   &0.00(0.12)&0.02(0.09)&0.99(0.11))\\
%         & 0.5   &0.51(0.13)&0.48(0.10)&1.05(0.11)\\
%         &1      &1.03(0.12)&0.99(0.10)&1.06(0.11)\\
%         \hline\\
%          & -1  &-1.01(0.20)&-0.96(0.16)&2.11(0.22) \\
%         & -0.5 &-0.48(0.18)&-0.54(0.15)&2.03(0.21)  \\
%$N(0,4)$ &   0  &-0.01(0.15)&0.02(0.12)&1.95(0.17) \\
%         & 0.5  &0.51(0.17)&0.49(0.14)&2.12(0.23)\\
%         &1     &1.05(0.17)&1.02(0.15)&1.95(0.20) \\
%         \hline \\
%\end{tabular}\\
%\end{table}

\begin{table}
\caption{Comparison of new estimator (${\hat\gamma}_G$) with Heckman's (${\hat\gamma}_H$) when individual effects are distributed as $0.5N(-6,9)+0.5N(6,9)$ ($200$ replicates with sample size $n=3000$).} \label{table4}\centering
\begin{tabular}{c c c c c}
\hline\\
    &$\gamma$& RMSE(${\hat\gamma}_G$)& RMSE(${\hat\gamma}_H$)&RMSE(${\hat\sigma}_H$)\\
        \hline\\
         & -1  &  0.37  & 0.81   &3.81\\
         & -0.5&0.29&    0.75  & 3.82 \\
        & 0  & 0.30  &  0.64&   3.86\\
         & 0.5 &  0.29   & 0.59 &   3.81\\
         &1  & 0.30 &   0.53  & 3.85\\
         \hline\\
         \end{tabular}\\
\end{table}

%\begin{table}
%\caption{Comparison of new estimator (${\hat\gamma}_G$) with Heckman's (${\hat\gamma}_H$) when individual effects are distributed as $0.5N(-6,9)+0.5N(6,9)$ ($200$ replicates with sample size %$n=3000$).} \label{table4}\centering
%\begin{tabular}{c c c c c}
%\hline\\
%    &$\gamma$& ${\hat\gamma}_G$(se)& ${\hat\gamma}_H$(se)&${\hat\sigma}_H$(se)\\
%        \hline\\
%         & -1  &   -1.10(0.35)  & -0.20(0.14)   &10.49(0.47)\\
%         & -0.5&  -0.52 (0.29)&    0.24(0.15)  &10.50 (0.46) \\
%        & 0  &  -0.05(0.29)  &  0.62(0.16)&   10.54(0.46))\\
%         & 0.5 &  0.51(0.29 )  & 1.06(0.19) &  10.49(0.44)\\
%         &1  & 1.00(0.30) &  1.48( 0.22 )  & 10.53(0.45)\\
%         \hline\\
%         \end{tabular}\\
%\end{table}

\section{An example}

We analyze the data set listed in Table (\ref{table5}) which has previously been considered by Heckman(1981). The dynamics of female labor supply is investigated based on panel data from the years 1968 to 1970, and 1971 to 1973. Model ({\ref{LM}}) is applied to estimate the dynamic parameter with $T=3$ and $x_{it}\equiv0$. Let $n_{ijl}$ be the number of observations of runs pattern $(i,j,l)$ in Table (\ref{table5}) for $i,j,l=0,1$.  As in Section 2.2, the following methods can be developed to estimate $\gamma$. The estimates are given in Table (\ref{table6}), where ${\hat\gamma}_G$ is the new estimator and ${\hat\gamma}_H$ and ${\hat\sigma}_H$ are Heckman's estimators.
$$
\displaystyle\hat{\gamma}_G=\arg\max\{p_{001}^{n_{001}}p_{010}^{n_{010}}p_{100}^{n_{100}}p_{110}^{n_{110}}p_{011}^{n_{011}}p_{101}^{n_{101}}\}
$$
where
$$
p_{001}=\frac{\int \Phi(-t)\Phi(-t)\Phi(t)dt}{K_1},p_{010}=\frac{\int \Phi(-t)\Phi(t)\Phi(-t-\gamma)dt}{K_1},p_{100}=\frac{\int\Phi(t)\Phi(-t-\gamma)\Phi(-t)dt}{K_1},
$$
$$
p_{110}=\frac{\int \Phi(t)\Phi(t+\gamma)\Phi(-t-\gamma)dt}{K_2},p_{011}=\frac{\int \Phi(-t)\Phi(t)\Phi(t+\gamma)dt}{K_2},p_{101}=\frac{\int\Phi(t)\Phi(-t-\gamma)\Phi(t)dt}{K_2},
$$
and
$$
K_1={\int \Phi(-t)\Phi(-t)\Phi(t)dt+\int \Phi(t)\Phi(-t-\gamma)\Phi(-t)dt+\int \Phi(t)\Phi(-t-\gamma)\Phi(-t)dt},
$$
$$
K_2=\int \Phi(t)\Phi(t+\gamma)\Phi(-t-\gamma)dt+\int \Phi(-t)\Phi(t)\Phi(t+\gamma)dt+\int\Phi(t)\Phi(-t-\gamma)\Phi(t)dt.
$$

\par

From the analyzed results in the age group 49-59 and runs pattern from 1971 to 1973, neither Heckman's method nor the proposed method yield evidence of a dynamic relationship, and perhaps more data needs to be collected. However, the difference for the older group between the period 1968-170 and 1971-1973 is significant; the difference for the younger group between the period 1968-170 and 1971-1973 is not significant.
%However, the difference between the period 1968-170 and 1971-1973 for this age group is significant???\marginpar{\scriptsize can we find out if significant? that could be of interest}
For age group 30-44, both the proposed method and Heckman's method yield a significant dynamic relationship, with a positive estimated value of $\gamma$ %{\color{red}
 (here, positivity of $\gamma$ implies the unsurprising result that currently holding a job increases the likelihood of holding a job in future).%}\marginpar{\scriptsize put this in brackets}

%by p-values, $1-\Phi(|-0.163|/0.259)=0.265$ and $1-\Phi(|-0.275|/0.363)=0.224$.
%Maybe they will retire soon and there is no strong desire to hold a job. For women not old, the analyzed results show there has significant dynamic relation by Heckman's methods and the proposed method. Women who once held a job is more likely to work that women who did not have work experience.

%Although not significantly so, $\hat\gamma_G$ is negative for women in the age group 49-59, and is significantly smaller than $\hat\gamma_G$ for women in the age group 30-44.  This indicates that women in the older age group are more likely to become unemployed than women in the younger age group.

\begin{table}
\caption{Runs patterns in the data ($1$ corresponds to work in the year, $0$ corresponds to no work)\label{table5}}\centering
\begin{tabular}{c c c S c c c S}
\hline\\
\multicolumn{3}{c}{Runs patterns}& {No. of}&\multicolumn{3}{c}{Runs pattern}& {No.of}\\
1968&1969&1970& {observations}&1971&1972&1973&{observations}\\
\hline\\
&\multicolumn{7}{c}{women aged 45-59 in 1968}\\[3mm]
0&0&0&87&0&0&0&96\\
0&0&1&5&0&0&1&5\\
0&1&0&5&0&1&0&4\\
1&0&0&4&1&0&0&8\\
1&1&0&8&1&1&0&5\\
0&1&1&10&0&1&1&2\\
1&0&1&1&1&0&1&2\\
1&1&1&78&1&1&1&76\\[3mm]
&\multicolumn{7}{c}{women aged 30-44 in 1968}\\[3mm]
0&0&0&126&0&0&0&133\\
0&0&1&16&0&0&1&13\\
0&1&0&4&0&1&0&5\\
1&0&0&12&1&0&0&16\\
1&1&0&24&1&1&0&8\\
0&1&1&20&0&1&1&19\\
1&0&1&5&1&0&1&8\\
1&1&1&125&1&1&1&130\\
\hline\end{tabular}
\end{table}

\begin{table}
\caption{Comparison of new estimator (${\hat\gamma}_G$) with Heckman's (${\hat\gamma}_H$) for data in Table \ref{table5} \label{table6}}\centering
\begin{tabular}{  r r r r r r}
\hline\\
\multicolumn{3}{c}{panel data(1969-1970)}&\multicolumn{3}{c}{panel data(1971-1973)}\\
 \hline\\
  {${\hat\gamma}_{G}$ (s.e.)} &  {${\hat\gamma}_H$  (s.e.)} & {${\hat\sigma}_H$  (s.e.)} & {${\hat\gamma}_{G}$  (s.e.)} &  {${\hat\gamma}_H$ (s.e.)} & {${\hat\sigma}_H$ (s.e.)}\\
 \hline \\

\multicolumn{6}{c}{ women aged 45-59 in 1968}  \\[2mm]
0.62{ (0.20)} & 0.54{ (0.27)} & 3.24{ (0.65)}&$-0.16${ (0.26)}&$-0.28${ (0.36)}&5.59{ (1.33)}\\\\
\multicolumn{6}{c}{women aged 30-44 in 1968}\\[2mm]
0.48{ (0.13)} & 0.47{ (0.17)} & 2.15{ (0.28)}&0.51{ (0.14)}&0.43{ (0.19)}& 2.63{ (0.37)}\\
\hline
\end{tabular}
\end{table}
%\marginpar{\scriptsize for table 6, I put 2d.p.\ which is enough I think and easier to read}

%\section{An Example}

\newpage

\section*{ Appendix}

{\bf Lemma 1.} If $f(x)$ satisfies the conditions given in Theorem 1, then
$$
\int\Phi(x)\Phi(-x-\gamma)f(x)dx=f(\mu_{\tau})\int\Phi(x)\Phi(-x-\gamma)dx+o(\sigma^{-1}_{\tau})
$$
and
 $$
 \int\Phi(-x)\Phi(x)f(x)dx=f(\mu_\tau)\int\Phi(-x)\Phi(x)dx+o(\sigma^{-1}_{\tau}).
 $$
\par
{\bf Proof.}
\begin{eqnarray*}
&&\left|\sigma_\tau\left[\int\Phi(x)\Phi(-x-\gamma)f(x)dx-f(\mu_\tau)\int\Phi(x)\Phi(-x-\gamma)dx\right]\right|\\ \\
&&=\left|\int\Phi(x)\Phi(-x-\gamma)h(\frac{x-\mu_\tau}{\sigma_\tau})dx-h(0)\int\Phi(x)\Phi(-x-\gamma)dx\right|\\ \\
&&\leq \int_{x>M}\Phi(x)\Phi(-x-\gamma)h(\frac{x-\mu_\tau}{\sigma_\tau})dx+\int_{x<-M}\Phi(x)\Phi(-x-\gamma)h(\frac{x-\mu_\tau}{\sigma_\tau})dx\\ \\
&&\;+h(0)\int_{x>M}\Phi(x)\Phi(-x-\gamma)dx+h(0)\int_{x<-M}\Phi(x)\Phi(-x-\gamma)dx\\ \\
&&\;+\int_{|x|\leq M}\Phi(x)\Phi(-x-\gamma)\left|h(\frac{x-\mu_\tau}{\sigma_\tau})-h(0) \right|dx\\ \\
&&\leq \Phi(-M-\gamma)+\Phi(-M)+h(0)\int_{x>M}\Phi(x)\Phi(-x-\gamma)dx\\ \\
&&\;\;\;+h(0)\int_{x<-M}\Phi(x)\Phi(-x-\gamma)dx+\int_{|x|\leq M}\Phi(x)\Phi(-x-\gamma)\left|h(\frac{x-\mu_\tau}{\sigma_\tau})-h(0) \right|dx.
\end{eqnarray*}
\par
For given $\gamma$ , $\Phi(-M-\gamma)$ and $\Phi(-M)$ can be arbitrary small for sufficient large $M$. Furthermore $\int\Phi(x)\Phi(-x-\gamma)$ is integrable, and so
$\int_{x<-M}\Phi(x)\Phi(-x-\gamma)dx$ and $\int_{x>M}\Phi(x)\Phi(-x-\gamma)dx$ can also be arbitrary small for sufficient large $M$. For given $M$, $\int_{|x|\leq M}\Phi(x)\Phi(-x-\gamma)\left|h(\frac{x-\mu_\tau}{\sigma_\tau})-h(0) \right|dx$ can also be arbitrary small for sufficient large $\sigma_\tau$. So
$$
\int\Phi(x)\Phi(-x-\gamma)f(x)dx=f(\mu_\tau)\int\Phi(x)\Phi(-x-\gamma)dx+o(\sigma^{-1}_{\tau}).
$$
Similarly, the other part can be proved.
\vskip1em

\vskip1em
\par
{\bf Lemma 2.}
$$
\int\Phi(-x)\Phi(x+\bbeta)dx=\bbeta\Phi(\frac{\bbeta}{\sqrt{2}})+\frac{1}{\sqrt{\pi}}\exp\{-\frac{\bbeta^2}{4}\}.
$$
\par
{\bf Proof.} By the fact $d(x\Phi(x)+\phi(x))=\Phi(x)$ and integration by parts,
\begin{eqnarray*}
\int\Phi(-x)\Phi(x+\bbeta)dx&=&\int\phi(x)[(x+\bbeta)\Phi(x+\bbeta)+\phi(x+\bbeta)]dx\\
&=&\bbeta\int\phi(x)\Phi(x+\bbeta)dx+\int x\phi(x)\Phi(x+\bbeta)dx+\int\phi(x)\phi(x+\bbeta)dx \\
&=&\bbeta\Phi(\frac{\bbeta}{\sqrt{2}})+2\int\phi(x)\phi(x+\bbeta)dx\\
&=&\bbeta\Phi(\frac{\bbeta}{\sqrt{2}})+\frac{1}{\sqrt{\pi}}\exp\{-\frac{\bbeta^2}{4}\}.
\end{eqnarray*}
\vskip1em

{\bf Lemma 3.} Suppose $\sigma_\tau=a\sqrt{n}$($a>0$) and then
$$
\frac{1}{n^{1/4}}\begin{pmatrix}
\sum\limits_{i=1}^{n}\left[I_{\{d_{i1}=1,d_{i2}=0\}}-EI_{\{d_{i1}=1,d_{i2}=0\}}\right]\\ \\
\sum\limits_{i=1}^{n}\left[I_{\{d_{i1}=0,d_{i2}=1\}}-EI_{\{d_{i1}=0,d_{i2}=1\}}\right]
\end{pmatrix}
\stackrel{d}\longrightarrow N(0,\Sigma)
$$
where
$$
\Sigma=\frac{h(0)}{a}\begin{pmatrix}
\int\Phi(x)\Phi(-x-\gamma)dx& &0\\ \\
0&& \int\Phi(x)\Phi(-x)dx
\end{pmatrix}.
$$
\par
{\bf Proof:} For $c_1, c_2\in R$, let
$$
U_{i\:n}=c_1\left[I_{\{d_{i1}=1,d_{i2}=0\}}-EI_{\{d_{i1}=1,d_{i2}=0\}}\right]+c_2\left[I_{\{d_{i1}=0,d_{i2}=1\}}-EI_{\{d_{i1}=0,d_{i2}=1\}}\right]
$$
and then
$$
E(U_{i\:n})=0,\;\;\;\sqrt{n}E(U^2_{i\:n})= \frac{h(0)}{a}\left[c_1^2\int\Phi(x)\Phi(-x-\gamma)dx+c_2^2 \int\Phi(x)\Phi(-x)dx\right]+o(1).
$$
By simple computations,
\begin{eqnarray*}
E[\exp\{U_{i\:n}t/n^{1/4}\}]&=&1+\frac{t^2}{2\sqrt{n}}E(U_{i\:t}^2)+E[o(\frac{U^2_{i\:n}}{n^{1/2}})]\\
&=&1+\frac{t^2}{2\sqrt{n}}E(U_{i\:t}^2)+o(n^{-1})\\
&=&1+\frac{ h(0)\left[c_1^2\int\Phi(x)\Phi(-x-\gamma)dx+c_2^2 \int\Phi(x)\Phi(-x)dx\right]t^2}{2an}+o(n^{-1}).
\end{eqnarray*}

The moment generating function of $\sum\limits_{i=1}^{n}U_{i\:n}/n^{1/4}$ is
\begin{eqnarray*}
\phi_n(t)&=&E[\exp\{\sum\limits_{i=1}^{n}U_{i\:n}t/n^{1/4}\}]\\
&=&[E(\exp\{U_{i\:n}t/n^{1/4}\})]^n\\
&=&\left\{1+\frac{ h(0)\left[c_1^2\int\Phi(x)\Phi(-x-\gamma)dx+c_2^2 \int\Phi(x)\Phi(-x)dx\right]t^2}{2an}+o(n^{-1})
\right\}^n\\
&\longrightarrow & \exp\{\frac{a h(0)\left[c_1^2\int\Phi(x)\Phi(-x-\gamma)dx+c_2^2 \int\Phi(x)\Phi(-x)dx\right]t^2}{2a}\}
\end{eqnarray*}
which implies the Lemma holds.

\vskip1em
\par

{\bf Lemma 4. }
Suppose $\sigma_\tau=a\sqrt{n}$($a>0$) and the first derivative of $h(x)$ is continuous, and then
$$
n^{1/4}\begin{pmatrix}
\frac{\sum\limits_{i=1}^{n}I_{\{d_{i1}=1,d_{i2}=0\}}}{\sqrt{n}}-\frac{h(0)}{a}\int\Phi(x)\Phi(-x-\gamma)dx\\ \\
\frac{\sum\limits_{i=1}^{n}I_{\{d_{i1}=0,d_{i2}=1\}}}{\sqrt{n}}-\frac{h(0)}{a}\int\Phi(x)\Phi(-x)dx
\end{pmatrix}
\stackrel{d}\longrightarrow N(0,\Sigma)
$$
where
$$
\Sigma=\frac{h(0)}{a}\begin{pmatrix}
\int\Phi(x)\Phi(-x-\gamma)dx& &0\\ \\
0&& \int\Phi(x)\Phi(-x)dx
\end{pmatrix}.
$$

\par
{\bf Proof:} Since the first derivative of $h(x)$ is continuous and $\sigma_\tau=a\sqrt{n}$, we have
\begin{eqnarray*}
\sqrt{n}\times E I_{\{d_{11}=1,d_{12}=0\}}&=&\sqrt{n}\times\int \Phi(x)\Phi(-x-\gamma)f(x)dx\\
&=&\sqrt{n}\times\int\Phi(x)\Phi(-x-\gamma)\frac{1}{\sigma_\tau}h(\frac{x-\mu_\tau}{\sigma_\tau})dx\\
&=&\sqrt{n}\times\int\Phi(x)\Phi(-x-\gamma)\frac{1}{a\sqrt{n}}h(\frac{x-\mu_\tau}{\sigma_\tau})dx\\
&=&\frac{1}{a}\int\Phi(x)\Phi(-x-\gamma)h(\frac{x-\mu_\tau}{\sigma_\tau})dx\\
&=&\frac{h(0)}{a}\int\Phi(x)\Phi(-x-\gamma)dx+O(\sigma_\tau^{-1})\\
&=&\frac{h(0)}{a}\int\Phi(x)\Phi(-x-\gamma)dx+O(n^{-1/2}).
\end{eqnarray*}
Similarly, we can obtain
$$
\sqrt{n}\times E I_{\{d_{11}=0,d_{12}=1\}}=\frac{h(0)}{a}\int\Phi(x)\Phi(-x)dx+O(n^{-1/2}).
$$
\par
\begin{eqnarray*}
&&n^{1/4}\begin{pmatrix}
\frac{\sum\limits_{i=1}^{n}I_{\{d_{i1}=1,d_{i2}=0\}}}{\sqrt{n}}-\frac{h(0)}{a}\int\Phi(x)\Phi(-x-\gamma)dx\\ \\
\frac{\sum\limits_{i=1}^{n}I_{\{d_{i1}=0,d_{i2}=1\}}}{\sqrt{n}}-\frac{h(0)}{a}\int\Phi(x)\Phi(-x)dx
\end{pmatrix}\\
&&=n^{1/4}\begin{pmatrix}
\frac{\sum\limits_{i=1}^{n}[I_{\{d_{i1}=1,d_{i2}=0\}}-E I_{\{d_{i1}=1,d_{i2}=0\}}]}{\sqrt{n}}+\sqrt{n} E I_{\{d_{11}=1,d_{12}=0\}}-\frac{h(0)}{a}\int\Phi(x)\Phi(-x-\gamma)dx\\ \\
\frac{\sum\limits_{i=1}^{n}[I_{\{d_{i1}=0,d_{i2}=1\}}-E I_{\{d_{i1}=0,d_{i2}=1\}}]}{\sqrt{n}}+\sqrt{n} E I_{\{d_{11}=0,d_{12}=1\}}-\frac{h(0)}{a}\int\Phi(x)\Phi(-x)dx
\end{pmatrix}\\
&&=n^{1/4}\begin{pmatrix}
\frac{\sum\limits_{i=1}^{n}[I_{\{d_{i1}=1,d_{i2}=0\}}-E I_{\{d_{i1}=1,d_{i2}=0\}}]}{\sqrt{n}}\\ \\
\frac{\sum\limits_{i=1}^{n}[I_{\{d_{i1}=0,d_{i2}=1\}}-E I_{\{d_{i1}=0,d_{i2}=1\}}]}{\sqrt{n}}
\end{pmatrix}
+n^{1/4}\begin{pmatrix}
\sqrt{n} E I_{\{d_{11}=1,d_{12}=0\}}-\frac{h(0)}{a}\int\Phi(x)\Phi(-x-\gamma)dx\\ \\
+\sqrt{n} E I_{\{d_{11}=0,d_{12}=1\}}-\frac{h(0)}{a}\int\Phi(x)\Phi(-x)dx
\end{pmatrix}\\
&&=n^{-1/4}\begin{pmatrix}
\sum\limits_{i=1}^{n}[I_{\{d_{i1}=1,d_{i2}=0\}}-E I_{\{d_{i1}=1,d_{i2}=0\}}]\\ \\
\sum\limits_{i=1}^{n}[I_{\{d_{i1}=0,d_{i2}=1\}}-E I_{\{d_{i1}=0,d_{i2}=1\}}]
\end{pmatrix}+o(1)\\
\end{eqnarray*}
which implies the Lemma holds by Lemma 3.

	\par
	{\bf Proof Theorem 1}. (i) follows immediately from the law of large numbers,
Proposition 1 and continuity of $G(x)$.

To prove (ii), it follows from the delta method and Lemma 4 above that
	\begin{eqnarray*}
	&&n^{1/4}\left(W-G(\gamma)\right)\\ \\
	&&=n^{1/4}\left(\frac{\sum\limits_{i=1}^{n}I_{\{d_{i1}=1,d_{i2}=0\}}/\sqrt{n}}
	{\sum\limits_{i=1}^{n}I_{\{d_{i1}=0,d_{i2}=1\}}/\sqrt{n}}-\frac{\frac{h(0)}{a}\int\Phi(x)\Phi(-x-\gamma)dx}{\frac{h(0)}{a}\int\Phi(x)\Phi(-x)dx}
	\right)\\ \\
	&&\stackrel{d}\longrightarrow N(0,\sigma^{*2})
	\end{eqnarray*}
	where
	$$
	\sigma^{*2}=\frac{a\int\Phi(x)\Phi(-x-\gamma)dx}{h(0)[\int\Phi(x)\Phi(-x)dx]^2}+\frac{a[\int\Phi(x)\Phi(-x-\gamma)dx]^2}{h(0)[\int\Phi(x)\Phi(-x)dx]^3}
	$$
	and then
	$$
	n^{1/4}\left(\hat{\gamma}-\gamma\right)=n^{1/4}\left(G^{-1}(W)-G^{-1}(G(\gamma))\right)\stackrel{d}\longrightarrow N(0,\frac{\sigma^{*2}}{[G^{'}(\gamma)]^2}).
	$$

	So
	$$
	\sqrt{ \sum\limits_{i=1}^{n}I_{\{d_{i1}=0,\;d_{i2}=1\}}
	}\left(\hat{\gamma}-\gamma\right)\stackrel{d}\longrightarrow N(0,\sigma^2)
	$$
	by
	$$
	 \frac{\sum\limits_{i=1}^{n}I_{\{d_{i1}=0,\;d_{i2}=1\}}}{\sqrt{n}}\stackrel{p}\longrightarrow \frac{h(0)\int\Phi(x)\Phi(-x)dx}{a}.
	$$
	\vskip1em

\vskip1em
\par
{\bf Lemma 5.} Let ${\bx}_1,{\bx}_2,\cdots,{\bx}_k,{\bx}_{k+1}\in R^{k}$ satisfy: (a) ${\bx}_1,{\bx}_2,\cdots,{\bx}_k$ are linearly independent;
(b) ${\bx}_{k+1}=-c_1{\bx}_1-c_2{\bx}_2-\cdots-c_k{\bx}_k$ where $c_1,\cdots,c_k$ are non-negative real number, and $r_1,\cdots,r_k,r_{k+1}$ be positive real number, then the equation
\begin{equation} \left\{
   \begin{array}{c}
     G({\bx}_1^{'}{\bbeta}+\alpha)-r_1G(-{\bx}_1^{'}{\bbeta}) =0\\ \\
    G({\bx}_2^{'}{\bbeta}+\alpha)-r_2G(-{\bx}_2^{'}{\bbeta}) =0\\ \\
    \cdots\cdots \\ \\
    G({\bx}_k^{'}{\bbeta}+\alpha)-r_kG(-{\bx}_k^{'}{\bbeta}) =0\\ \\
     G({\bx}_{p+1}^{'}{\bbeta}+\alpha)-r_{k+1}G(-{\bx}_{k+1}^{'}{\bbeta}) =0
   \end{array}
 \right.\label{E1}\end{equation}
has a unique solution $\bbeta$ and $\alpha$.
\par
{\bf Proof:} For fixed $\alpha$, let
$$
u_\alpha(z)=\frac{G(z+\alpha)}{G(-z)}
$$
and
\begin{eqnarray*}
\frac{du_\alpha(z)}{dz}&=&\frac{G^{'}(z+\alpha)G(-z)+G(z+\alpha)G^{'}(-z)}{G^2(-z)}\\
                      &=&-\sqrt{\pi}\frac{\Phi(-(z+\alpha)/\sqrt{2})G(-z)+G(z+\alpha)\Phi(z/\sqrt{2})}{G^2(-z)}\\
                      &<&0.
\end{eqnarray*}
So $u_\alpha(z)$ is decreasing in $z$ and $\displaystyle\lim_{z\rightarrow-\infty}u_\alpha(z)=\infty$ and $\displaystyle\lim_{z\rightarrow\infty}u_\alpha(z)=0$. Thus for fixed $\alpha$, the equation
\begin{equation} \left\{
   \begin{array}{c}
     G({\bx}_1^{'}{\bbeta}+\alpha)-r_1G(-{\bx}_1^{'}{\bbeta}) =0\\ \\
    G({\bx}_2^{'}{\bbeta}+\alpha)-r_2G(-{\bx}_2^{'}{\bbeta}) =0\\ \\
    \cdots\cdots \\ \\
    G({\bx}_k^{'}{\bbeta}+\alpha)-r_kG(-{\bx}_k^{'}{\bbeta}) =0
\end{array}
 \right.\label{E2}\end{equation}
 has a unique solution when ${\bx}_1,\cdots,{\bx}_k$ are linearly independent.
 \par
 Let $\bbeta^{*}=(\bbeta_1(\alpha),\cdots,\bbeta_k(\alpha))^{'}$ the solution of (\ref{E2}), and then
 $$
\frac{d\bbeta^{*}}{d \alpha}=-{X^{'}}^{-1}\delta
$$
where
$$
\delta=(\delta_1,\cdots,\delta_k)^{'},\;\;\;\delta_i=\frac{\Phi(-({\bx}_i^{'}\bbeta^{*}+\alpha)/\sqrt{2})}{\Phi(-({\bx}_i^{'}\bbeta^{*}+\alpha)/\sqrt{2})
+r_i\Phi({\bx}_i^{'}\bbeta^{*}/\sqrt{2})}
$$
and
$$
X=({\bx}_1,{\bx}_2,\cdots,{\bx}_k).
$$
Define
$$
t(\alpha)=G({\bx}_{k+1}^{'}\bbeta^{*}+\alpha)-r_{k+1}G(-{\bx}_{k+1}^{'}\bbeta^{*}),
$$
 and then
\begin{eqnarray*}
 \frac{dt(\alpha)}{d\alpha}&=&-\sqrt{\pi}\left\{ \left[\Phi(-\frac{{\bx}_{k+1}^{'}\bbeta^{*}+\alpha}{\sqrt{2}})+r_{k+1}\Phi(\frac{{\bx}_{k+1}^{'}\bbeta^{*}}{\sqrt{2}})
 \right]{\bx}_{k+1}^{'}
\frac{d\bbeta^{*}}{d\alpha}+\Phi(-\frac{{\bx}_{k+1}^{'}\bbeta^{*}+\alpha}{\sqrt{2}})\right\}\\
&=&-\sqrt{\pi}\left\{ \left[\Phi(-\frac{{\bx}_{k+1}^{'}\bbeta^{*}+\alpha}{\sqrt{2}})+r_{k+1}\Phi(\frac{{\bx}_{k+1}^{'}\bbeta^{*}}{\sqrt{2}})
 \right]\left(\sum\limits_{j=1}^{k}c_j\delta_i\right)+\Phi(-\frac{{\bx}_{k+1}^{'}\bbeta^{*}+\alpha}{\sqrt{2}})\right\}\\
 &<& 0,
\end{eqnarray*}
 which implies $t(\alpha)=0$ has an unique solution and the lemma is proved.

\par
\par
{\bf Proof of Theorem 2}. By Lemma 5 given in the above, it can be proved with $r_i=p_i/(1-p_i)$ and ${\bx}_i={\bx}_{i2}-{\bx}_{i1}$.

\par
{\bf Proof of Corollary}. Without loss of generality, suppose that ${\bx}_{12}-{\bx}_{11},\cdots,{\bx}_{k2}-{\bx}_{k1}$ are linearly independent and
$$
{\bx}_{k+1\;2}-{\bx}_{k+1\;1}=a_1({\bx}_{1 2}-{\bx}_{1 1})+\cdots+a_k({\bx}_{k 2}-{\bx}_{k 1})
$$
where $a_1,\cdots,a_k$ is a non-positive real number. Then the determinant
\begin{equation}
\begin{vmatrix}
{\bx}^{'}_{12}-{\bx}^{'}_{11}&1\\
{\bx}^{'}_{22}-{\bx}^{'}_{21}&1\\
\vdots&\vdots\\
{\bx}^{'}_{k2}-{\bx}^{'}_{k1}&1\\
{\bx}^{'}_{k+1\;2}-{\bx}^{'}_{k+1\;1}&1
\end{vmatrix}
\label{PMA}
\end{equation}
is equal to
\begin{eqnarray*}
&&\begin{vmatrix}
{\bx}^{'}_{12}-{\bx}^{'}_{11}\\
{\bx}^{'}_{22}-{\bx}^{'}_{21}\\
\vdots\\
{\bx}^{'}_{k2}-{\bx}^{'}_{k1}
\end{vmatrix}
\left[1-\left({\bx}_{k+1\;2}-{\bx}_{k+1\;1}\right)^{'}
\begin{pmatrix}
{\bx}^{'}_{12}-{\bx}^{'}_{11}\\
{\bx}^{'}_{22}-{\bx}^{'}_{21}\\
\vdots\\
{\bx}^{'}_{k2}-{\bx}^{'}_{k1}
\end{pmatrix}
^{-1}{\bf 1}_k\right]\\
&&= \begin{vmatrix}
{\bx}_{12}-{\bx}_{11},{\bx}_{22}-{\bx}_{21},\cdots, {\bx}_{k2}-{\bx}_{k1}\end{vmatrix}\left[1-\sum\limits_{i=1}^{k}a_i\right]\neq 0
\end{eqnarray*}
by the assumption. This implies that the rank of (\ref{PMA}) is $k+1$.
\par
Since the rank of $({\bf 1}_m ,{\bX^{*}}^{'})$ is equal to that of $({\bX^{*}}^{'},{\bf 1}_m )$, which is a $m\times(k+1)$ matrix,  and (\ref{PMA}) is a matrix obtained by the first $k+1$ rows of $({\bX^{*}}^{'},{\bf 1}_m )$, thus the rank of $({\bf 1}_m,{\bX^{*}}^{'} )$
is $k+1$.

\begin{description}
{\centerline{ \large {\underline {References}}}}

\item Arellano, M. (2003). Discrete choices with panel data. {\sl Investigaciones Economicas}, 27, 423-458.

\item Arellano, M. and Bonhomme, S. (2009). Robust priors in nonlinear panel data models. {\sl Econometrica}, 77, 489-536.

\item{} Arellano, M. and Honore, B. (2001). Panel data Models: some recent developments. {\sl Handbook of Econometrics}, Vol. V, ed. by J. Heckman and E. Leamer. Amsterdam: North Holland.

\item  Bartolucci, F. and Farcomeni, A.(2009). A multivariate extension of the dynamic logit model for longitudinal data based on a latent Markov heterogeneity structure. {\sl Journal of the American Statistical Association}, 104, 816-833.

 \item Bartolucci, F. and Nigro, V.(2010). A dynamic model for binary panel data with unobserved heterogeneity admitting a $\sqrt{n}$ consistent conditional estimator. {\sl Econometrica}, 78, 719-733.

 \item Chamberlain, G.(1980). Analysis of covariance with qualitative data. {\sl Review of Economic Studies}, 47, 225-238.

  \item Chamberlain, G.(1985). Heterogeneity, omitted variables bias, and duration dependence. {\sl Longitudinal Analysis of Labor Market Data}, edited by Heckman, J. and Singer, B. Cambridge University Press.

\item{} Heckman, J.(1978). Simple statistical models for discrete panel data developed and applied to test the hypothesis of true state dependence against the hypothesis of spurious state dependence. {\sl Annales de l'lNSEE} 30/31, 227-269.

\item Heckman, J.(1980). The incidental parameters problem and the problem of initial conditions in estimating a discrete time-discrete data stochastic process. {\sl Structural Analysis of Discrete Data with Econometric Applications}, ed by C. F. Manski and D. McFadden, p179-195. Cambridge, MA: MIT Press.
 \item Heckman, J.(1981). Heterogeneity and state dependence. {\sl Studies in Labor Markets}, ed by S. Rosen, p91-140. University of Chicago Press.

\item Hisao, C.(2003). {\sl Analysis of Panel Data}(Second Ed.). New York: Cambridge University Press.

\item Honore, B. and Kyriazidou, E.(2000). Panel data discrete choice models with lagged dependent variables. {\sl Econometrica}, 68, 611-629.

\item Horowitz, J. L.(1992). A smoothed maximum score estimator for binary response model. {\sl Econmetrica}, 60, 505-531.

\item{} Lancaster, T.(2000). The incidental parameter problem since 1984. {\sl Journal of Econometrics}, 95, 391-413.

\item{} Lancaster, T.(2002). Orthogonal parameters and panel data. {\sl Review of Economic Studies}, 647-666.

\item Manski, C. (1987). Semiparametric analysis of random effects linear models from binary panel data. {\sl Econometrica}, 55, 357-362.

\item McCullagh, P. and Nelder, J. A.(1989). {\sl Generalized Linear Models}. London: Chapman \& Hall.

\item Neyman, J. and Scott, E. S.(1948). Consistent estimation from partially consistent observations. {\sl Econometrica}, 16, 1-32.

\end{description}
\end{document}